\documentclass[aps,prc,reprint,groupedaddress]{revtex4-1}
\usepackage{graphicx} 
\usepackage{dcolumn} 
\usepackage{bm}
\usepackage{hyperref}

\begin{document}

\title{Nuclear fragmentation and charge-exchange reactions induced by pions in the $\Delta$-resonance region}
\author{Zhao-Qing Feng}
\email{Corresponding author: fengzhq@impcas.ac.cn}

\affiliation{Institute of Modern Physics, Chinese Academy of Sciences, Lanzhou 730000, People's Republic of China}

\date{\today}
\begin{abstract}
Dynamics of the nuclear fragmentations and the charge exchange reactions in pion-nucleus collisions near the $\Delta$(1232) resonance energies has been investigated within the Lanzhou quantum molecular dynamics (LQMD) transport model. An isospin, momentum and density-dependent pion-nucleon potential is implemented in the model, which influences the pion dynamics, in particular the kinetic energy spectra, but weakly impacts the fragmentation mechanism. The absorption process in pion-nucleon collisions to form the $\Delta$(1232) resonance dominates the heating mechanism of target nucleus. The excitation energy transferred to the target nucleus increases with the pion kinetic energy and is similar for both $\pi^{-}$ and $\pi^{+}$ induced reactions. The magnitude of fragmentation of target nucleus weakly depends on the pion energy. The isospin ratio in the pion double charge exchange is influenced by the isospin ingredient of target nucleus.

\begin{description}
\item[PACS number(s)]
25.80.Hp, 24.10.Lx, 21.65.Jk
\end{description}
\end{abstract}

\maketitle

\section{Introduction}

The pion dynamics is of importance in understanding the transportation process in heavy-ion collisions and in hadron induced reactions, which also contributes other particle production as the main reaction channels, such as eta, strange particles etc. Pion production in heavy-ion collisions near threshold energies has been investigated for extracting the high-density behavior of the nuclear symmetry energy (isospin asymmetric part of equation of state) \cite{Fe06,Xi09,Fe10a,Xi13}. The dynamics of pions produced in heavy-ion collisions is a complicated process, which is related to the pion-nucleon and $\Delta(1232)$-nucleon interactions, the decay of resonances etc. The in-medium properties of pion and $\Delta(1232)$ are not well understood up to now, which are related the isospin effects. The in-medium effects of pions in heavy-ion collisions have been studied in Refs \cite{Fe05,So15} for threshold energy corrections and in Refs \cite{Xi93,Fu97,Fe10b,Ho14} for the pion optical potential. The baryon density of pion produced varies with the evolution dynamics in nucleus-nucleus collisions, in which parts of pions are created above normal nuclear density. However, the pion-nucleus reactions have advantages to investigate the pion and $\Delta(1232)$ properties in nuclear medium around the saturation density. On the other hand, the energy deposition in the pion induced reactions would be helpful in exploring the nuclear explosion of highly excited nucleus.

In the past four decades, pion-nucleus collisions at the $\Delta$-resonance energies have been extensively investigated within the meson factories, i.e., the Los Alamos Meson Physics Facility (LAMPF) in New Mexico, United States \cite{Ka79,Ka83,An86,Ro92,Fo07}; the Paul Scherrer Institute (PSI) in Villigen, Switzerland \cite{Mi80}; the Tri-University Meson Facility (TRIUMF) in Vancouver, Canada \cite{Ie15}; and superconducting kaon spectrometer (SKS) at KEK 12-GeV PS, Japan \cite{Kr05}. Several theoretical approaches were developed for understanding the pion induced nuclear reactions \cite{En94,Bu06,Le02}. The energy released by pion-nucleon collisions heats the target nucleus. Consequently, fast-nucleon emission, particle evaporation, intermediate mass fragments, fission etc will proceed. A number of nuclides are produced tending to the line of $\beta$-stability \cite{Ka79}. In this work, a microscopic transport approach is used to describe the pion-nucleus collisions in the $\Delta$-resonance energies. The charge-exchange reactions and fragmentation mechanism of target nucleus are particularly concentrated on.

\section{Brief description of the model}

The Lanzhou quantum molecular dynamics (LQMD) model is used to investigate the nuclear dynamics induced by pions in the $\Delta$-resonance region for the first time. The isospin physics, particle production, in-medium properties of hadrons, (hyper-)fragment production etc in heavy-ion collisions and antiproton (proton) induced reactions have been investigated within the model \cite{Fe11,Fe12,Fe14}. I implemented an isospin, density and momentum dependent mean-field potentials based on the Skyrme forces for nucleons and resonances. The optical potentials for hyperons and kaons in nuclear medium are derived from the effective Lagrangian approaches. In the model, all possible reaction channels in hadron-hadron collisions were included, i.e., charge-exchange reactions, elastic and inelastic collisions by distinguishing isospin effects, annihilation reactions in collisions of antiparticles and particles etc. The temporal evolutions of the hadrons under the self-consistently generated mean-field potentials are governed by Hamilton's equations of motion.

The pion dynamics is influenced by the mean-field potential in nuclear medium. The potential is composed of the Coulomb interaction between the charged particles and pion-nucleon potential. The optical potential is evaluated from the in-medium energy of pions $V_{\pi}^{opt}(\textbf{p}_{i},\rho_{i}) = \omega_{\pi}(\textbf{p}_{i},\rho_{i}) - \sqrt{m_{\pi}^{2}+\textbf{p}_{i}^{2}}$. The isoscalar and isovector contributions to the pion energy are included as follows \cite{Fe15}
\begin{equation}
\omega_{\pi}(\textbf{p}_{i},\rho_{i}) = \omega_{isoscalar}(\textbf{p}_{i},\rho_{i})+C_{\pi}\tau_{z}\delta (\rho/\rho_{0})^{\gamma_{\pi}}.
\end{equation}
The coefficient $C_{\pi}= \rho_{0} \hbar^{3}/(4f^{2}_{\pi}) = 36$ MeV is taken from fitting the experimental data of pion-nucleus scattering. The isospin quantities are taken to being $\tau_{z}=$ 1, 0, and -1 for $\pi^{-}$, $\pi^{0}$ and $\pi^{+}$, respectively. The isospin asymmetry $\delta=(\rho_{n}-\rho_{p})/(\rho_{n}+\rho_{p})$ and the quantity $\gamma_{\pi}$ adjusts the stiffness of isospin splitting of the pion-nucleon potential. We take the $\gamma_{p} = 2$ in the work. The isoscalar part of the pion self-energy in the nuclear medium is evaluated via the $\Delta$-hole model. Shown in Fig. 1 is the optical potential as functions of baryon density and pion momentum, respectively. It should be noticed that the potential reaches the minimum at the $\Delta$ resonance energy, i.e., $E_{lab}=$0.19 GeV or $p=$0.298 GeV/c, which dominates the pion dynamics and enhances the pion absorption in nuclear medium.

\begin{figure*}
\includegraphics[width=16 cm]{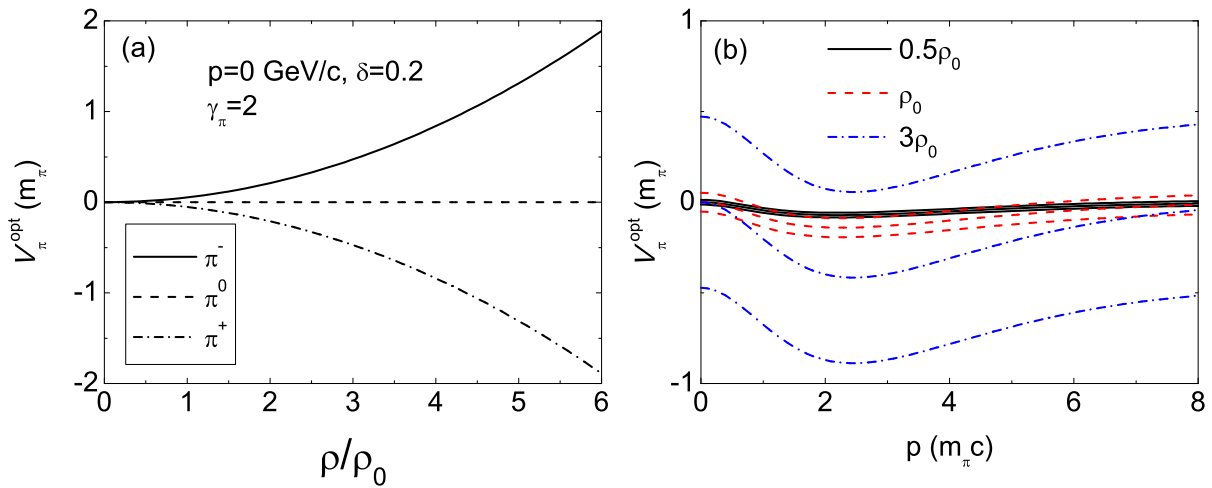}
\caption{\label{fig:wide} (Color online) Momentum and density dependence of the pion optical potential in dense nuclear matter with the isospin asymmetry of $\delta$=0.2.}
\end{figure*}

The probability in two-particle collisions to a channel is performed by using a Monte Carlo procedure via the relative distance smaller than the scattering radius. The channels associated with the pion induced reactions as follows:
\begin{eqnarray}
&& N\pi  \leftrightarrow \Delta,  \quad  N\pi \leftrightarrow N^{\ast}, \quad NN\pi (s-state) \leftrightarrow  NN, \nonumber \\
&& N\triangle \leftrightarrow NN, \quad  NN^{\ast} \leftrightarrow NN, \quad  \triangle\triangle \leftrightarrow NN.
\end{eqnarray}
The momentum-dependent decay widths are used for the resonances of $\Delta$(1232) and $N^{\ast}$(1440) \cite{Hu94}. We have taken a constant width of $\Gamma$=150 MeV for the $N^{\ast}$(1535) decay. The cross section of pion-nucleon scattering is evaluated with the Breit-Wigner formula as the form of
\begin{eqnarray}
\sigma_{\pi N\rightarrow R}(\sqrt{s}) = \sigma_{max}(|\textbf{p}_{0}/\textbf{p}|)^{2}\frac{0.25\Gamma^{2}(\textbf{p})}
{0.25\Gamma^{2}(\textbf{p})+(\sqrt{s}-m_{0})^{2}},
\end{eqnarray}
where the $\textbf{p}$ and $\textbf{p}_{0}$ are the momenta of pions at the energies of $\sqrt{s}$ and $m_{0}$, respectively, and $m_{0}$ being the centroid of resonance mass, e.g., 1.232 GeV, 1.44 GeV and 1.535 GeV for $\Delta$(1232), $N^{\ast}$(1440), and $N^{\ast}$(1535), respectively. The maximum cross section $\sigma_{max}$ is taken from fitting the total cross sections of the available experimental data in pion-nucleon scattering with the Breit-Wigner form of resonance formation \cite{Li01}. For example, 200 mb, 133.3 mb, and 66.7 mb for $\pi^{+}+p\rightarrow \Delta^{++}$ ($\pi^{-}+n\rightarrow \Delta^{-}$), $\pi^{0}+p\rightarrow \Delta^{+}$ ($\pi^{0}+n\rightarrow \Delta^{0}$) and $\pi^{-}+p\rightarrow \Delta^{0}$ ($\pi^{+}+n\rightarrow \Delta^{+}$), respectively. And 24 mb, 12 mb, 32 mb, 16 mb for $\pi^{-}+p\rightarrow N^{\ast 0}(1440)$ ($\pi^{+}+n\rightarrow N^{\ast +}(1440)$), $\pi^{0}+p\rightarrow N^{\ast +}(1440)$ ($\pi^{0}+n\rightarrow N^{\ast 0}(1440)$), $\pi^{-}+p\rightarrow N^{\ast 0}(1535)$ ($\pi^{+}+n\rightarrow N^{\ast +}(1535)$) and $\pi^{0}+p\rightarrow N^{\ast +}(1535)$ ($\pi^{0}+n\rightarrow N^{\ast 0}(1535)$), respectively. Shown in Fig. 2 is a comparison of the elastic and total cross sections in the pion-nucleon collisions with the available data from the PDG collaboration \cite{Ol14}. The total cross sections include the sum of three resonances and the contributions of strangeness production. It is obvious that the more resonances are needed to be implemented at the pion momentum above 0.5 GeV/c. All possible contributions of nucleonic resonances and strangeness resonances have been included in the Giessen Boltzmann Uehling Uhlenbeck (GiBUU) transport model \cite{Bu12}. The spectra can be reproduced nicely well with the resonance approach, in particular in the domain of $\Delta$ resonance momenta (0.298 GeV/c). The pion-nucleon scattering dominates the energy deposition in the pion-induced reactions, which contributes the fragmentation of target nuclei.

\begin{figure*}
\includegraphics[width=18 cm]{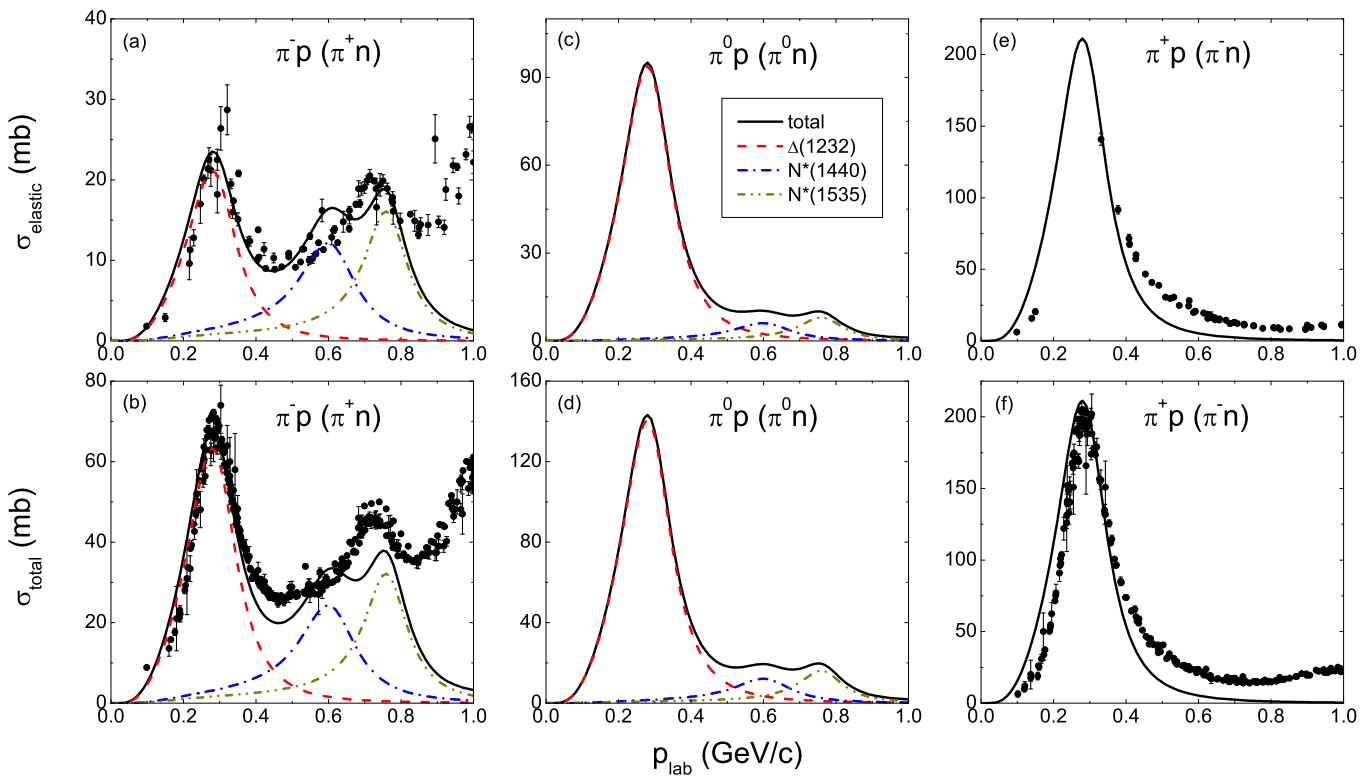}
\caption{\label{fig:wide} (Color online) The elastic (upper panel) and total (down panel) pion-nucleon scattering cross sections contributed from different resonances. The available data are taken from the PDG collaboration \cite{Ol14}. }
\end{figure*}

\section{Results and discussion}

Nuclear reactions induced by pions in the $\Delta$ resonance region provide the opportunity to study the pion-nucleon interaction, the charge-exchange reactions, and the in-medium properties of $\Delta$ resonance, which are not well understood up to now. Shown in Fig. 3 is the kinetic energy distributions of $\pi^{-}$, $\pi^{0}$ and $\pi^{+}$ produced in collisions of $\pi^{-}$ and $\pi^{+}$ on $^{40}$Ca at the $\Delta$ resonance energy, respectively. The double-charge exchange reactions $\pi^{\pm}A\rightarrow \pi^{\mp}X$ have non-negligible contributions on the pion production. The total multiplicities of $\pi^{-}$ and $\pi^{+}$ are 0.55 and 0.12 for the $\pi^{-}$ induced reactions, and being 0.05 and 0.69 for the $\pi^{+}$. The Coulomb interaction between charged pions and protons influences the absorption probability of pions in nuclear medium. The pion-nucleon potential changes the kinetic energy spectra, i.e., reducing the energetic pion production. But the total pion yields weakly depend on the potential. More pronounced effect from the potential is observed for a heavier target as shown in Fig. 4.

\begin{figure*}
\includegraphics[width=16 cm]{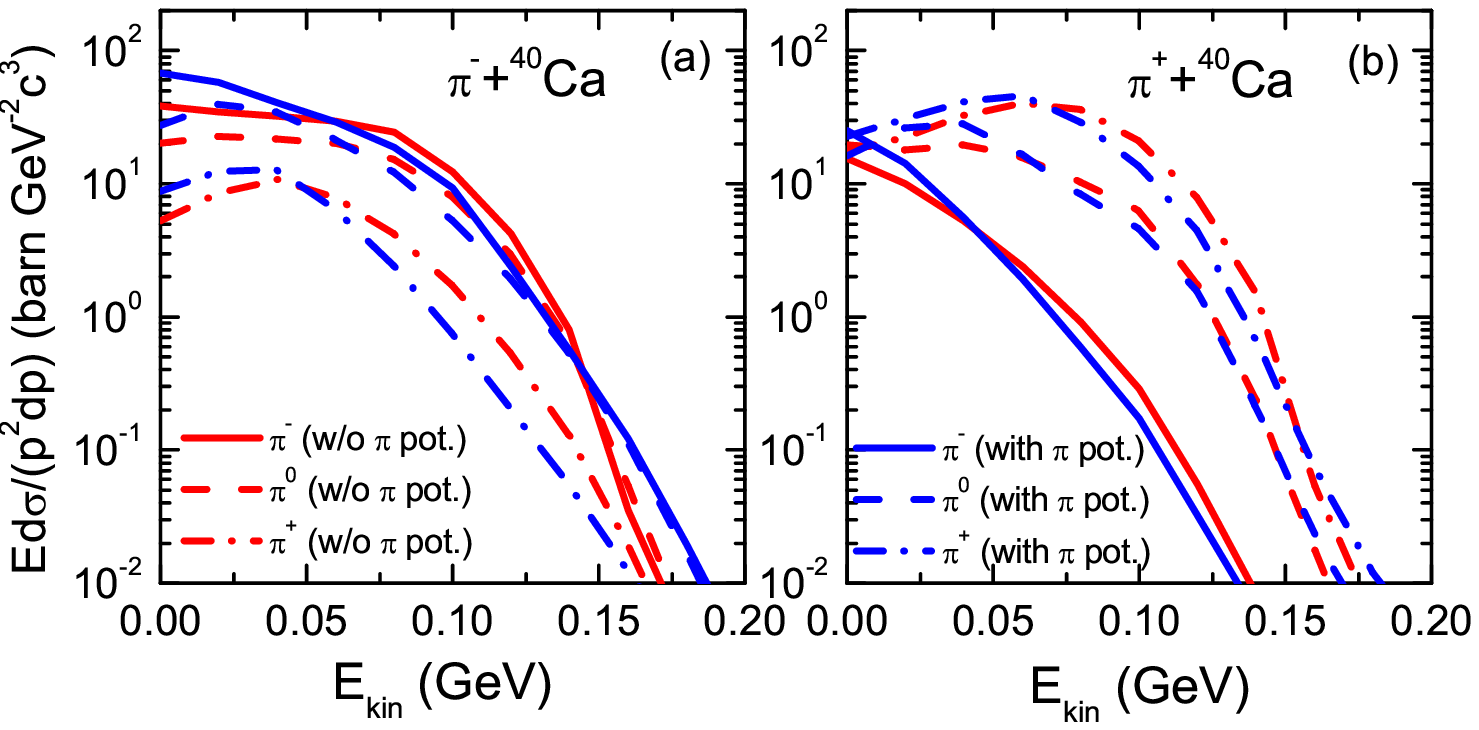}
\caption{\label{fig:wide} (Color online) Kinetic energy spectra of $\pi^{-}$, $\pi^{0}$ and $\pi^{+}$ produced in collisions of $\pi^{-}$ and $\pi^{+}$ on $^{40}$Ca at an incident momentum of 300 MeV/c without (red lines) and with (blue lines) the pion-nucleon potentials, respectively.}
\end{figure*}

\begin{figure*}
\includegraphics[width=16 cm]{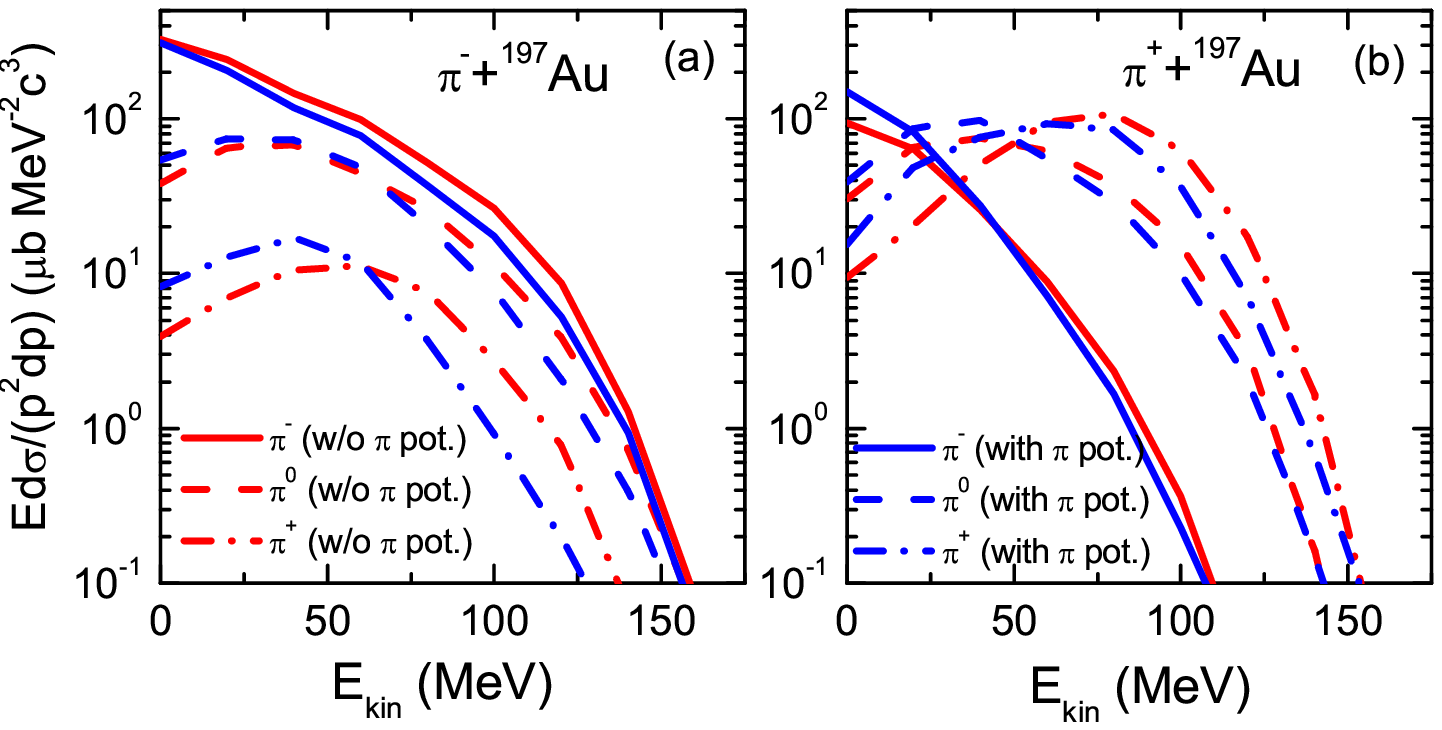}
\caption{\label{fig:wide} (Color online) The same as in Fig. 3, but for the target of $^{197}$Au at an incident kinetic energy of 180 MeV.}
\end{figure*}

In the past decades, the double charge exchange (DCX) in the pion induced reactions attracts much attention, which provides the possibility to explore the multiple pion interaction in a nucleus and is involved in two or more nucleons. Shown in Fig. 5 is the rapidity and transverse momentum distributions of pion production on the target of $^{40}$Ca at the incident momentum of 300 MeV/c. The structure of pion production from the DCX, single charge exchange and elastic scattering is very similar in phase space. The DCX cross sections are lower and depends on isospin ingredients of target nuclei, i.e., $R(\pi^{+}/\pi^{-})$ and $R(\pi^{-}/\pi^{+})$ being 0.22 and 0.07 for the reactions of ($\pi^{-}, \pi^{+}$) and ($\pi^{+}, \pi^{-}$) on $^{40}$Ca, respectively. The ratios are 0.07 and 0.13 for ($\pi^{-}, \pi^{+}$) and ($\pi^{+}, \pi^{-}$) on $^{197}$Au at the same energy. The process ($\pi^{-}, \pi^{+}$) can be understood via the reactions associated at least two protons, e.g., $\pi^{-}p\rightarrow\Delta^{0}$, $\Delta^{0}\rightarrow\pi^{0}n$, $\pi^{0}p\rightarrow\Delta^{+}$ and $\Delta^{+}\rightarrow\pi^{+}n$. In the neutron-rich nuclei, the process is constrained because of the larger collision probabilities between $\pi^{-}$ ($\pi^{0}$) and neutrons, which leads to the decrease of the $\pi^{+}/\pi^{-}$ ratio in the pion DCX reactions. The opposite processes take place for the DCX ($\pi^{+}, \pi^{-}$) in the neutron-rich nuclei. The kinetic energy spectra of the pion DCX are calculated as shown in Fig. 6. The consistent trend with the available data at the LAMPF \cite{Wo92} is found. The maximal cross sections move from the kinetic energy of 30 MeV for the DCX ($\pi^{+}, \pi^{-}$) to 50 MeV for ($\pi^{-}, \pi^{+}$).

\begin{figure*}
\includegraphics[width=16 cm]{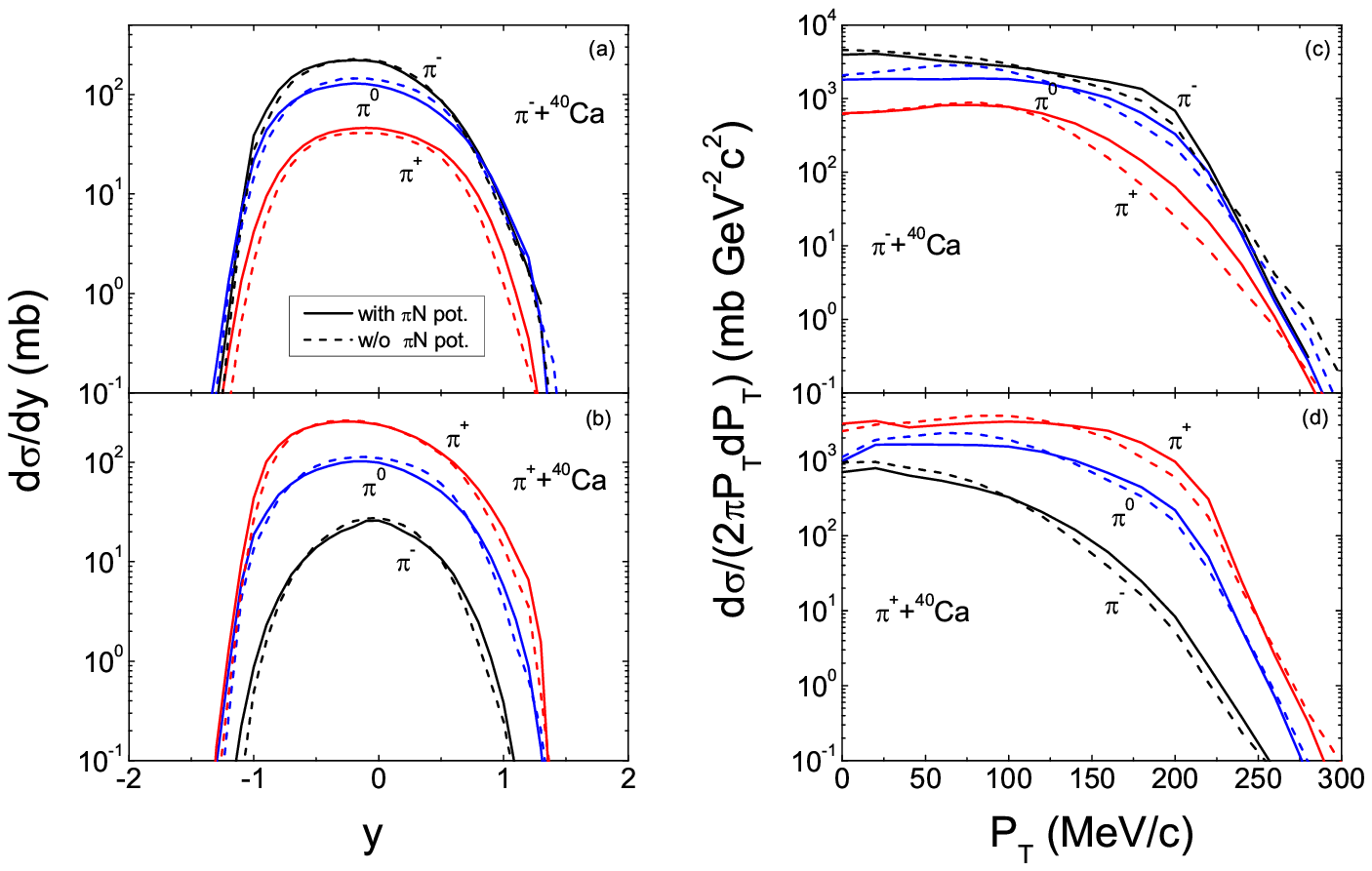}
\caption{\label{fig:wide} (Color online) Rapidity and transverse momentum distributions of $\pi^{-}$, $\pi^{0}$ and $\pi^{+}$ produced in pion induced reactions at the incident momentum of 300 MeV/c .}
\end{figure*}

\begin{figure*}
\includegraphics[width=16 cm]{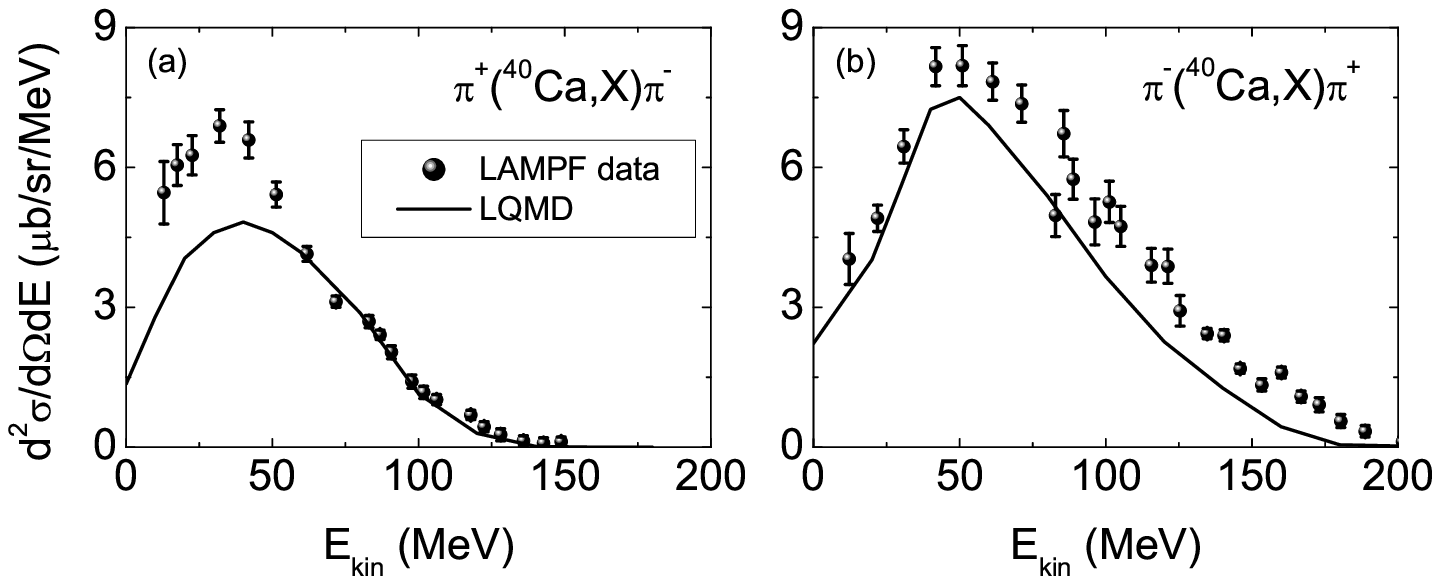}
\caption{\label{fig:wide} Kinetic energy spectra of pions at outgoing angle of 80$^{o}$ in double charge exchange reactions at incident energy of 180 MeV. The available data are from the LAMPF measurements \cite{Wo92}.}
\end{figure*}

The energy deposition in a nucleus by pion induced reactions is realized via the pion-nucleon collisions associated with the resonance production and reabsorption, which leads to the formation of highly excited nucleus. The fragmentation reactions induced by pions are described with the help of the LQMD transport model combined with the GEMINI statistical decay code for excited fragments \cite{Ch88}. The nuclear dynamics is described by the LQMD model. The primary fragments formed at freeze-out stage are constructed in phase space with a coalescence model, in which nucleons are considered to belong to one cluster with the relative momentum smaller than $P_{0}$ and with the relative distance smaller than $R_{0}$ (here $P_{0}$ = 200 MeV/c and $R_{0}$ = 3 fm). The freeze-out stage is assumed that the energy deposition is reached the equilibrium and the pions do not interact nucleons at the evolution of 300 fm/c taken in this work. The excitation energies of primary fragments are evaluated from the binding energy difference between the fragments and their ground-state ones. The de-excitation of the primary fragments leads to a broad mass distribution, in which the structure effects (shell correction, fission barrier, particle separation energy) contribute to the process. The phase-space distributions of fragments with charged number of $Z\geq$2 in collisions of charged pions on $^{40}$Ca is calculated as shown Fig. 7. It is interest that the pion-nucleon potential enhances the fragment production in the domain of high kinetic energy ($>$60 MeV), which is caused from the fact that the attractive potential increases the energy deposition, in particular for the $\pi^{+}$ induced reactions. Similar conclusions are found with the heavier target $^{197}$Au as shown Fig. 8.

\begin{figure*}
\includegraphics[width=16 cm]{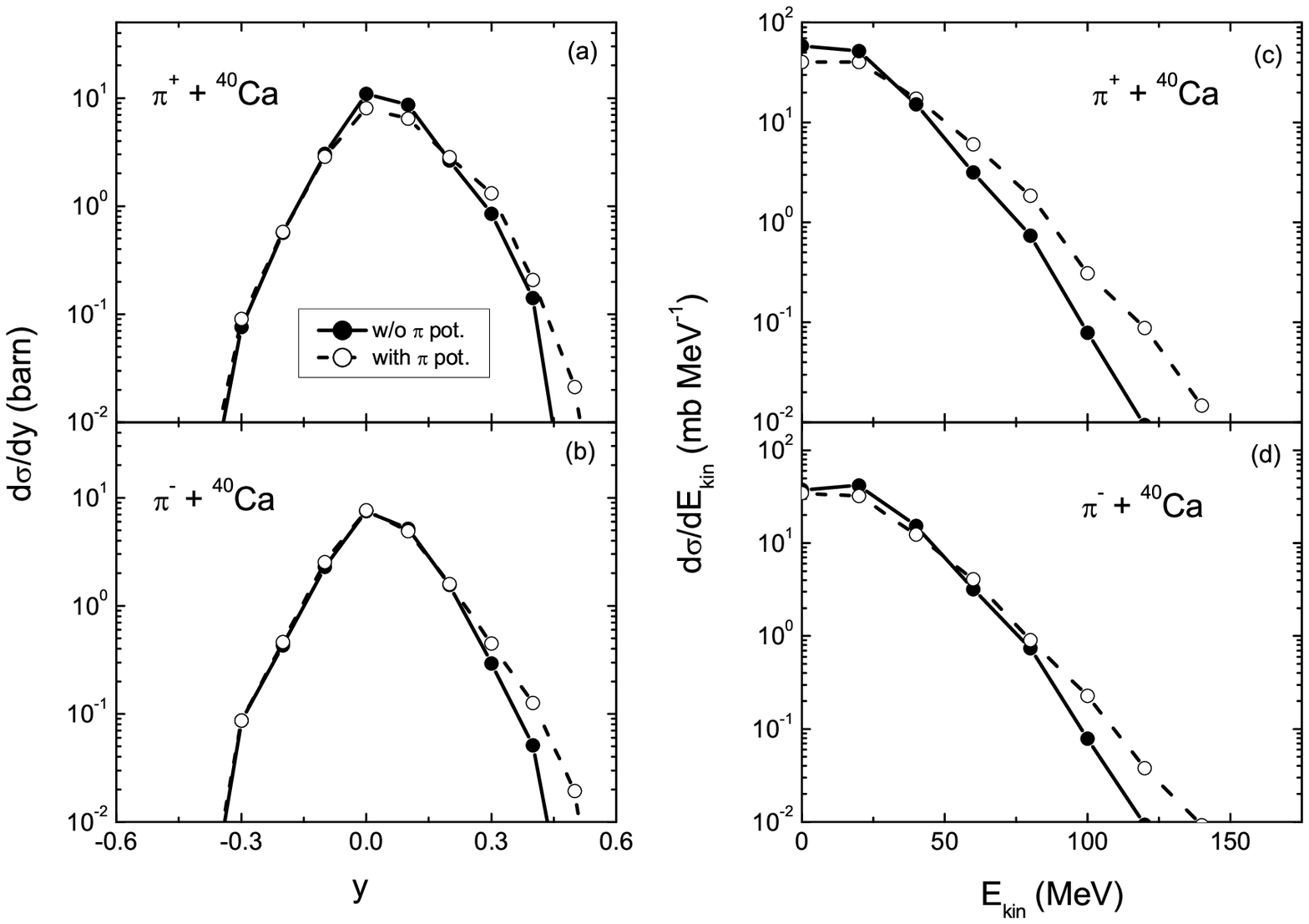}
\caption{\label{fig:wide} Rapidity and kinetic energy distributions of fragments with charged number of $Z\geq$2 in collisions of $\pi^{-}$ and $\pi^{+}$ on $^{40}$Ca at the incident momentum of 300 MeV/c, respectively.}
\end{figure*}

\begin{figure*}
\includegraphics[width=16 cm]{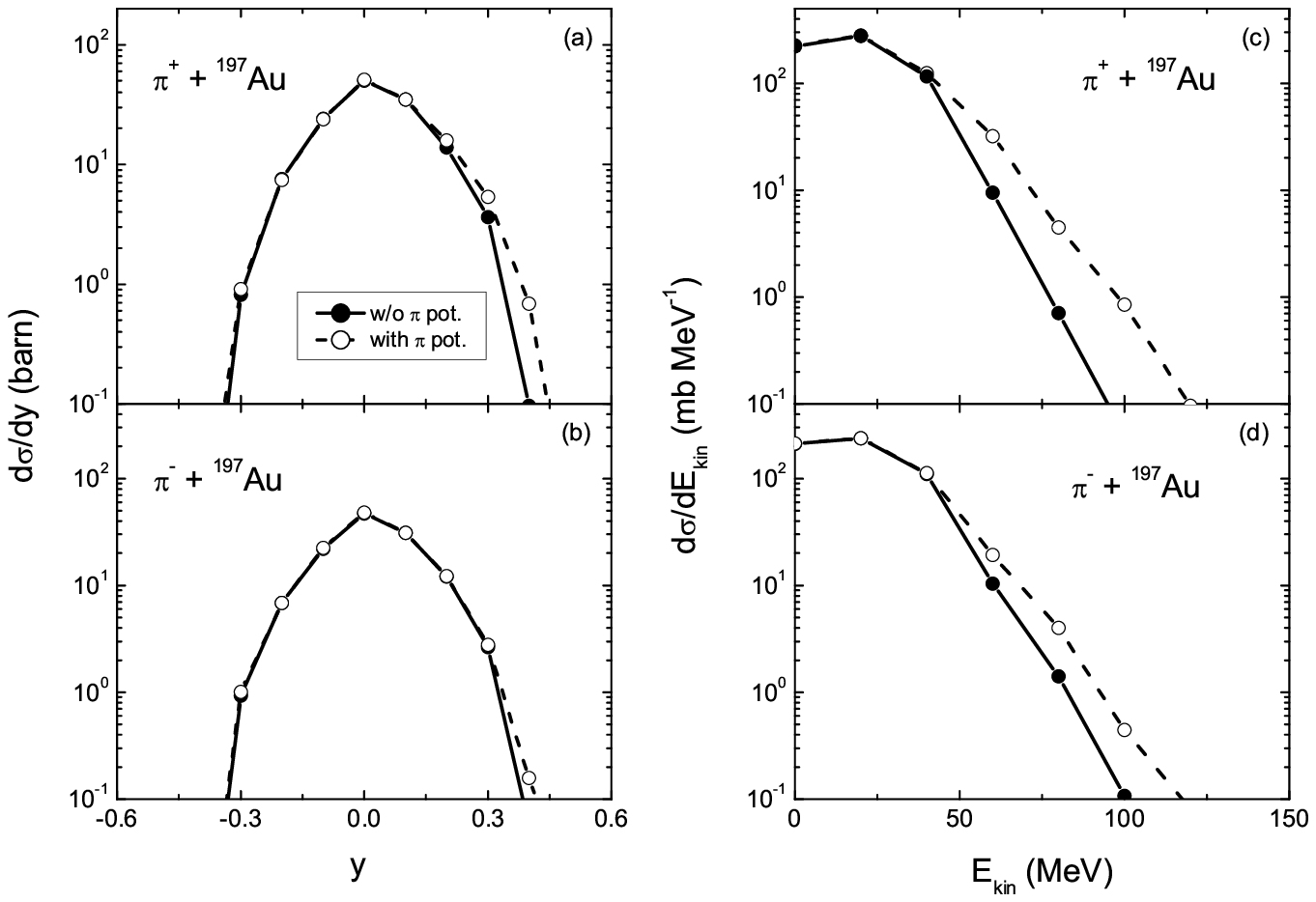}
\caption{\label{fig:wide} The same as in Fig. 5, but for the target of $^{197}$Au at an incident kinetic energy of 180 MeV.}
\end{figure*}

Fragmentation of target nucleus induced by pions is associated with the pion-nucleon scattering, $\Delta$ production and decay process, $\Delta$-nucleon interaction etc. The mechanism is related to the fast nucleon emission, particle evaporation after equilibrium, fission etc. Shown in Fig. 9 is the fragment distributions in $\pi^{-}$ and $\pi^{+}$ induced reactions on $^{40}$Ca at an incident momentum of 300 MeV/c. It is obvious that the pion-nucleon potential has negligible contribution on the fragment formation. The average mass removals are 4.8 and 4.7 for $\pi^{-}$ and $\pi^{+}$, respectively, which are evaluated from $\triangle A=A_{T}-\int_{A_{min}}^{A_{T}}\sigma(A)AdA/\int_{A_{min}}^{A_{T}}\sigma(A)dA$. Here, the $A_{T}$ and $A_{min}$ are the mass number of target nucleus and the integration limit being $A_{T}/2$. Comparison of calculations with the available experimental data from the LAMPF \cite{Ka79} is shown in Fig. 10 for the $\pi^{-}$ induced reactions and in Fig. 11 for bombarding $^{197}$Au with $\pi^{+}$ at the incident energy of 100 MeV, 180 MeV and 300 MeV, respectively. The fragment production in the target-mass region is nicely reproduced. The bump structure of intermediate mass fragments comes from the fission of heavy fragments. Similar mass and charge distributions are found for both $\pi^{-}$ and $\pi^{+}$ induced reactions.

\begin{figure*}
\includegraphics[width=16 cm]{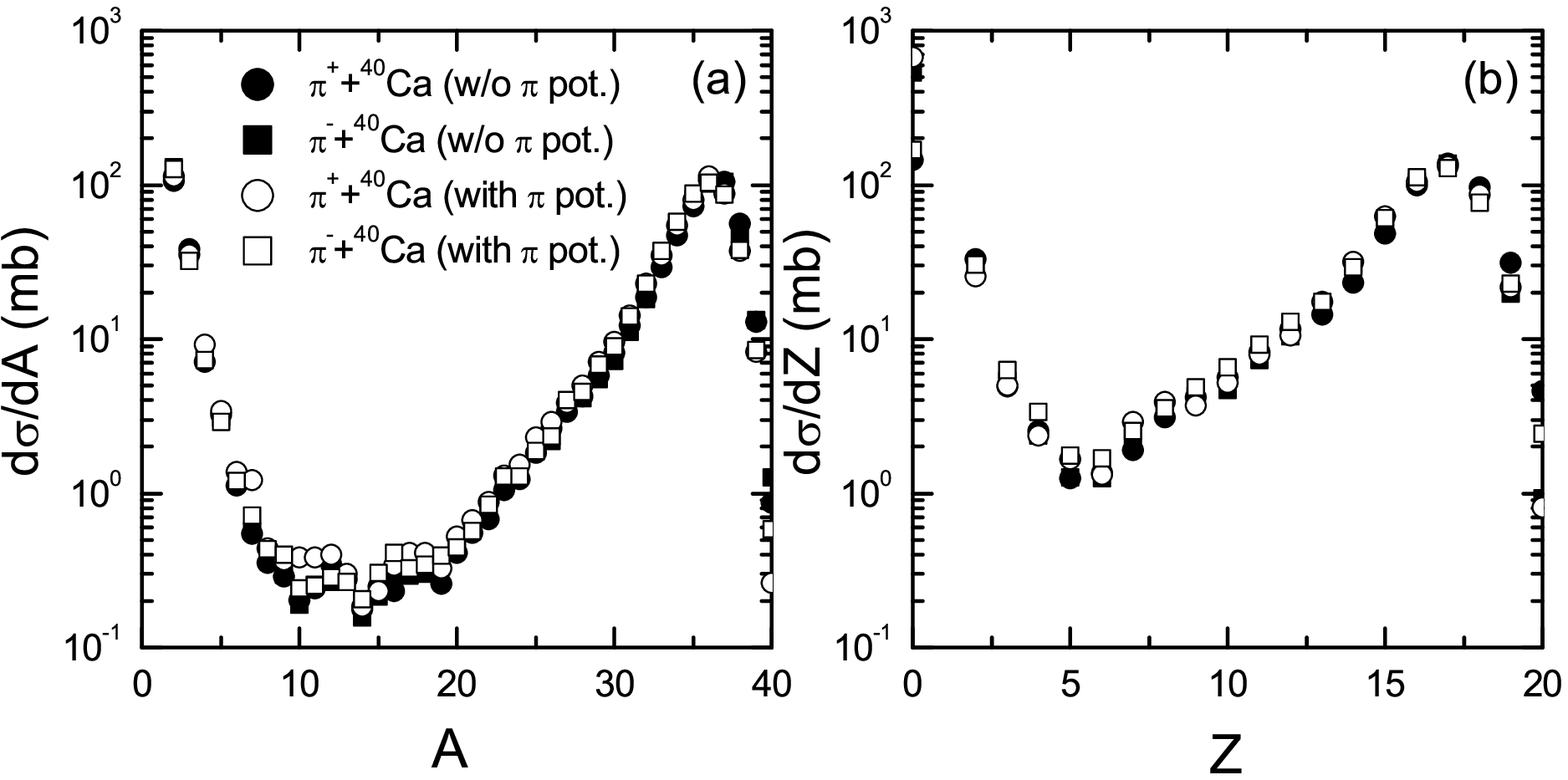}
\caption{\label{fig:wide} Mass and charge distributions of fragments produced in collisions of $\pi^{-}$ and $\pi^{+}$ on $^{40}$Ca at the incident momentum of 300 MeV/c, respectively.}
\end{figure*}

\begin{figure*}
\includegraphics[width=15 cm]{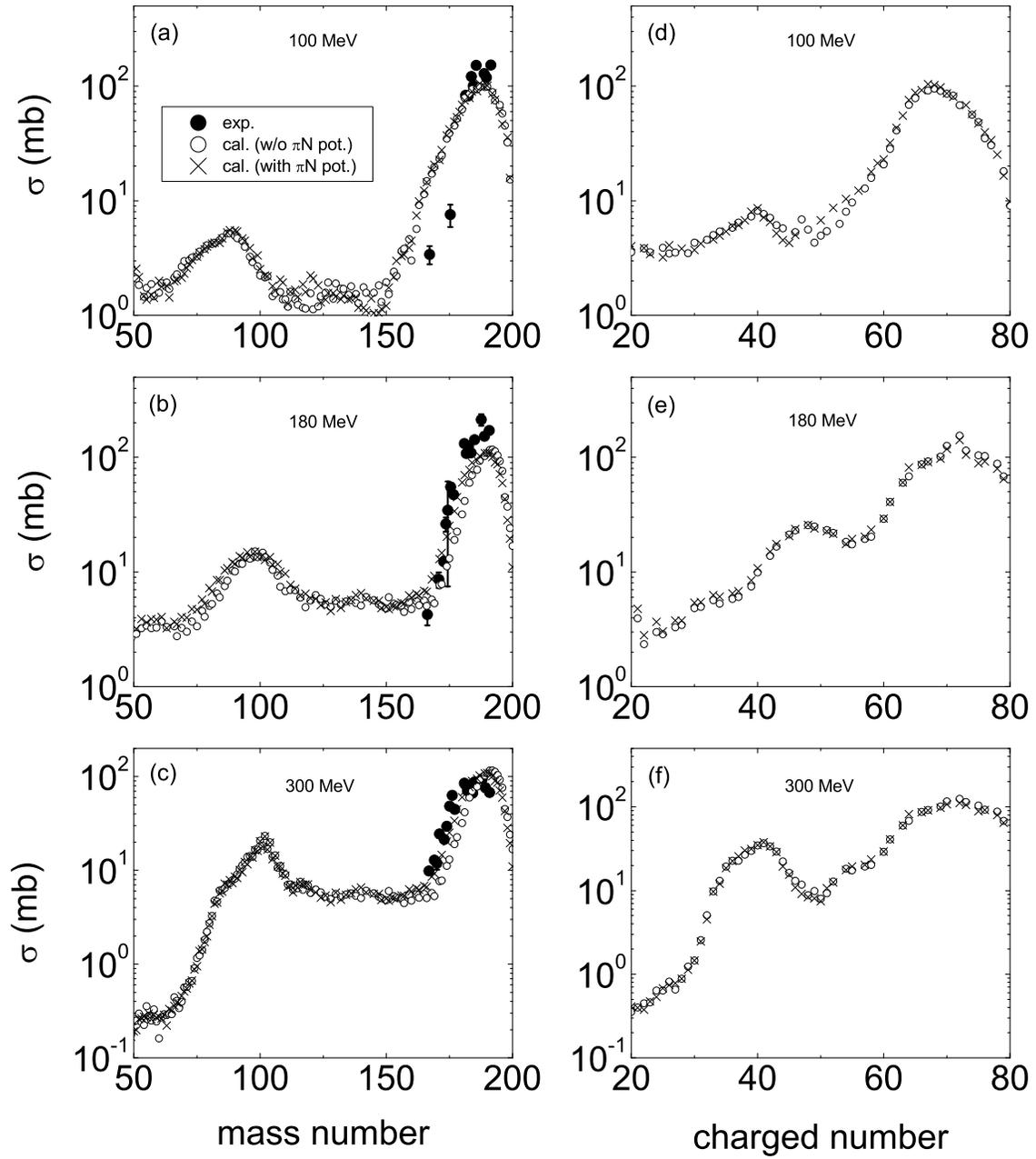}
\caption{\label{fig:wide} Fragment distributions with and without the pion-nucleon potential in the $\pi^{-}$+$^{197}$Au reaction at incident energy of 100 MeV, 180 MeV and 300 MeV, respectively. The available data are from the LAMPF measurements \cite{Ka79}. }
\end{figure*}

\begin{figure*}
\includegraphics[width=15 cm]{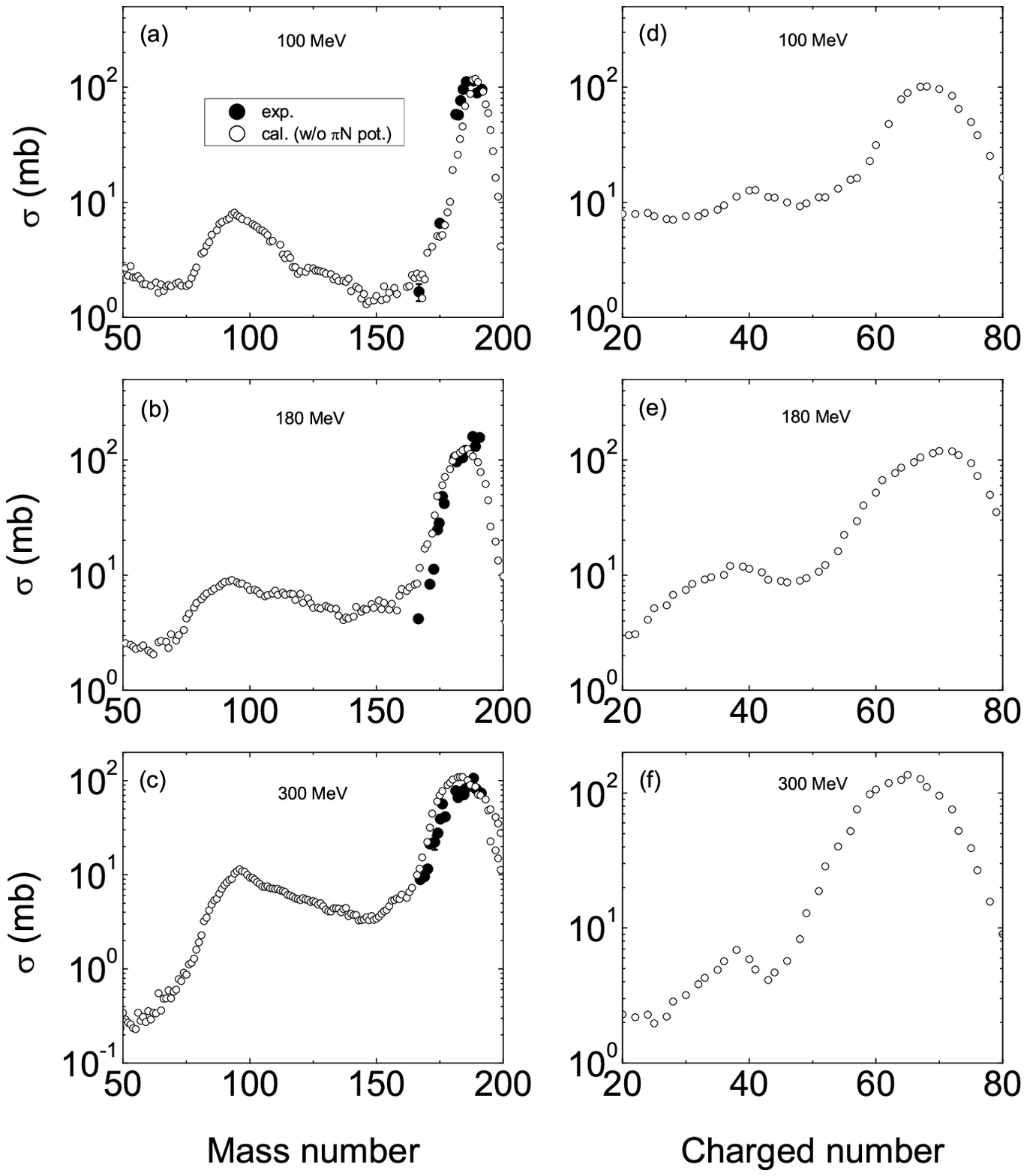}
\caption{\label{fig:wide} The same as in Fig. 8, but for the reaction of $\pi^{+}$+$^{197}$Au.}
\end{figure*}

Emission of the intermediate mass fragments (IMFs) is caused from the fluctuation of composite system, which has been investigated for extracting the nuclear equation of state and density dependence of symmetry energy in heavy-ion collisions \cite{Li04,Fe16}. I presented a comparison of IMFs correlated to charged particle multiplicity in pion induced reactions at different incident energies as shown Fig. 12. Production of the IMFs appears maximum near the charged particle multiplicity of 15. The IMF production is strongly suppressed in the pion induced reactions in comparison to heavy-ion collisions. It should be noticed that the shadow of fission products dominates the IMF production in the pion reactions. However, the fluctuation of nucleus-nucleus collisions leads to the multifragmentation of colliding partners in HICs, which contributes the IMF production. The fragmentation reactions induced by pions provide the basis in understanding the pion-nucleon interaction, energy deposition, fragment formation etc, which are helpful for investigating the hypernucleus production with the high-energy pion beams. The hypernucleus formation in pion-nucleus collisions is in progress.

\begin{figure*}
\includegraphics[width=16 cm]{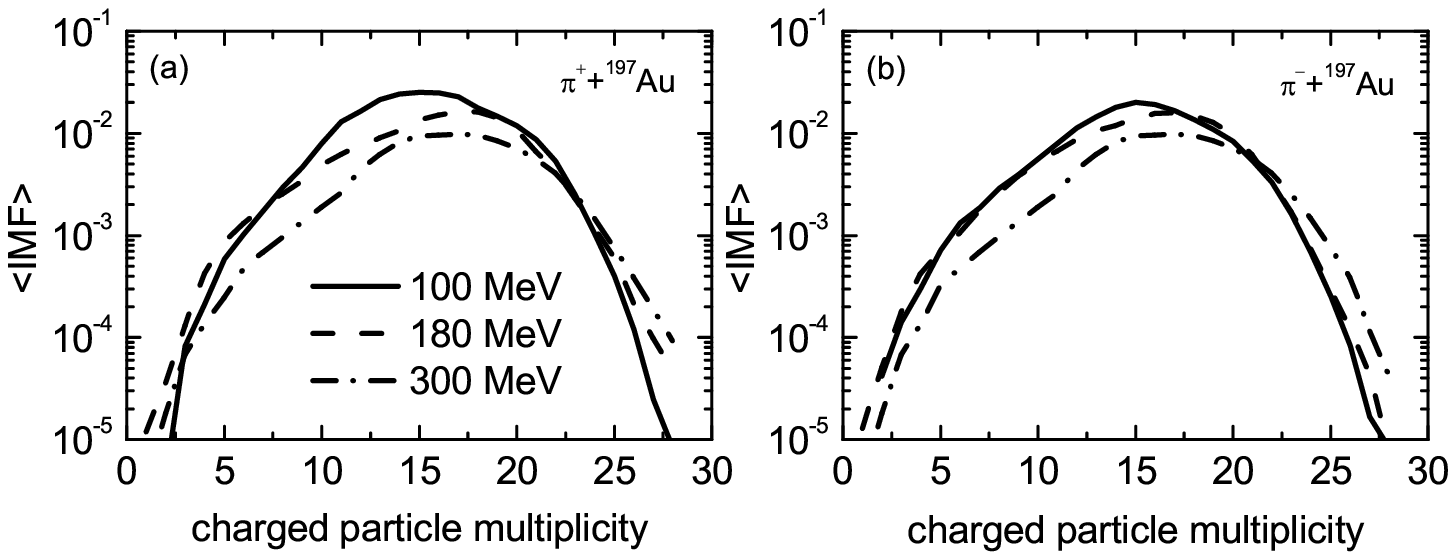}
\caption{\label{fig:wide} Correlation of the intermediate mass fragments (IMFs) and charged particle multiplicity in collisions of $\pi^{\pm}$+$^{197}$Au at incident energy of 100 MeV, 180 MeV and 300 MeV, respectively.}
\end{figure*}

\section{Conclusions}

The fragmentation mechanism and the charge-exchange reactions in pion induced nuclear reactions have been investigated within an isospin and momentum dependent hadron-transport model (LQMD). The pion-nucleon potential is of importance on the pion dynamics in the charge-exchange reactions. The isospin ratio in the pion DCX is influenced by the isospin ingredient of target nucleus. The attractive pion-nucleon potential near the $\Delta$-resonance energies (E=0.19 GeV, p=0.298 GeV/c) influences the kinetic energy spectra, but has negligible contribution on the fragmentation process. The relative motion energy is deposited in the nucleus via the pion-nucleon collisions. The transferred energy weakly depends on the incident pion energy. The pion induced reactions could be performed in the future facilities, such as the FAIR (GSI, Germany), HIAF (IMP, China) etc.

\bigskip
\textbf{Acknowledgements}

This work was supported by the Major State Basic Research Development Program in China (No. 2015CB856903), the National Natural Science Foundation of China Projects (Nos 11675226, 11175218 and U1332207), and the Youth Innovation Promotion Association of Chinese Academy of Sciences.

\end{document}